\documentclass[12pt]{iopart}

\usepackage{iopams}
\usepackage{enumerate}
\usepackage{bm}
\usepackage{graphicx,subfigure}

\begin{document}

\title{Scalar clouds around Kerr-Sen black holes}

\author{Yang Huang}
\address{Center for Astrophysics, Shanghai Normal University, 100 Guilin Road, Shanghai 200234, China}
\ead{saisehuang@163.com}
\author{Dao-Jun Liu}%
\address{Center for Astrophysics, Shanghai Normal University, 100 Guilin Road, Shanghai 200234, China}
\ead{djliu@shnu.edu.cn}
\author{Xiang-Hua Zhai}%
\address{Center for Astrophysics, Shanghai Normal University, 100 Guilin Road, Shanghai 200234, China}
\ead{zhaixh@shnu.edu.cn}
\author{Xin-Zhou Li}%
\address{Center for Astrophysics, Shanghai Normal University, 100 Guilin Road, Shanghai 200234, China}
\ead{kychz@shnu.edu.cn}

\begin{abstract}
In this paper, the behaviour of a charged massive scalar test field in the background of a Kerr-Sen black hole is investigated.  A kind of stationary solutions, dubbed scalar clouds, are obtained numerically and expressed by the existence lines in the parameter space. We show that for fixed background and a given set of harmonic indices, the mass and charge of the scalar clouds are limited in a finite region in the parameter space of the scalar field.  Particularly, the maximum values of the mass and charge of the clouds around extremal Kerr-Sen black holes are independent of the angular velocity of the black hole, whereas those in the extremal Kerr-Newman background depend on the angular velocity. In addition, it is demonstrated that, as the static limit of Kerr-Sen black hole,  Gibbons-Maeda-Garfinkle-Horowitz-Strominger black hole can not support scalar cloud.
\end{abstract}

\submitto{\CQG}
\maketitle

\section{Introduction}
Recently, much attention has been paid to scalar clouds around Kerr or Kerr-Newman black holes (BHs) \cite{PhysRevD.90.024051,Hod2013,PhysRevD.86.104026,PhysRevLett.112.221101,PhysRevD.90.104024}. These clouds are described by linearized stationary bound-state solutions with a real frequency
\begin{equation}\label{omega=omega_c}
	\omega=\omega_c,
\end{equation}
where $\omega_c$ is the critical frequency for superradiant scattering. Fully nonlinear counterparts of the linear clouds have also been shown to yield Kerr or Kerr-Newman BHs with scalar hair \cite{PhysRevLett.112.221101,PhysRevD.89.124018,doi:10.1142/S0218271814420140,Delgado2016234}. These hairy BHs may provide distinguishable phenomenology with future observations \cite{PhysRevLett.115.211102,Cunha2016}. Therefore, it is of interest to study the existence and  properties of the scalar clouds.

It has long been known that for scalar fields around a non-rotating BH, e.g., electrically charged Reissner-Nordstr\"{o}m (RN) BH, it is not possible to find a bound-state solution in the superradiant regime. In other words, there is no charged superradiant instability for asymptotically flat charged BHs \cite{PhysRevD.91.044047,Hod2012505,Hod20131489,Furuhashi01122004}. Accordingly, one can further conclude that there is no scalar clouds around   RN BHs, since bound states with frequency $\omega=\omega_c$ do not exist.

The formation of bound states of scalar field  requires that $\omega<\mu$, where $\mu$ denotes the mass of the scalar field. However, as is shown by Hod in a series of studies \cite{Hod2012320,0264-9381-32-13-134002,Hod2016181}, this condition
is not sufficient for massive scalar fields to form bound states in the background of Kerr spacetime and, the condition $\mu<\sqrt{2}\omega$ that is necessary for the existence of a potential well, should also be held. In a recent work \cite{PhysRevD.94.064030}, we find that the condition for the existence of bound-state solutions of a massive scalar field with charge $q$ in the background of a Kerr-Newman BH is given by
\begin{equation}\label{bound state condition}
	 f(\mu,q)\equiv\frac{qQ}{4M}+\sqrt{\frac{\mu^2}{2}+\frac{q^2Q^2}{16M^2}}<\omega<\mu,
\end{equation}
where $M$ and $Q$ are the mass and charge of the BH, respectively.

It is known that there are many exotic BH solutions in theories beyond Einstein-Maxwell theory.
Therefore, a natural and important question is whether the scalar clouds can form on other rotating BHs. The Kerr-Sen geometry, which is a four-dimensional rotating charged BH solution in the low energy limit of heterotic string field theory \cite{PhysRevLett.69.1006}, has attracted considerable attention in recent years. In fact, it has been found that charged massive scalar field is superradiantly unstable in the Kerr-Sen background \cite{S0218271815501023}.  By using an analytical method based on matching the asymptotic forms of the radial functions describing the scalar field, Bernard have shown that near-extremal Kerr-Sen BHs can support scalar clouds  \cite{PhysRevD.94.085007}.

The main goal of present paper is to perform a numerical study of charged massive scalar clouds around Kerr-Sen BHs beyond near-extremal limit. 
The paper is organized as follows. In Section \ref{Sec: background and equation of motion} we review briefly the  Kerr-Sen spacetime and the separation of variables for the scalar wave function as well as the boundary conditions for the radial equation. In Section \ref{Sec: effective potential},   the existence line of the scalar clouds is introduced and  the parameter space of the scalar field is analyzed with the help of three constraint lines. By the way, we demonstrate that the Gibbons-Maeda-Garfinkle-Horowitz-Strominger (GMGHS) BH, which is the static limit of the Kerr-Sen BH, can not support scalar clouds.  Then, our main numerical results are shown in Section \ref{Sec: existence lines}. Finally, we conclude the paper with a brief summary of our results. Throughout the paper, we use natural units in which $G=c=\hbar=1.$

\section{Charged massive  scalar field in Kerr-Sen Spactime}\label{Sec: background and equation of motion}
The Kerr-Sen spacetime, expressed  in the Boyer-Lindquist coordinates, is described by  the line element
\begin{eqnarray}
		ds^2&=&-\frac{\Delta}{\rho^2}\left(dt-a\sin^2\theta d\phi\right)^2+\rho^2\left(\frac{dr^2}{\Delta}+d\theta^2\right)\nonumber\\
		&&+\frac{\sin^2\theta}{\rho^2}\left[\left(r^2+2br+a^2\right)d\phi-adt\right]^2,
\end{eqnarray}
where
\begin{equation}
\rho^2=r^2+2br+a^2\cos^2\theta,\quad \Delta=r^2-2(M-b)r+a^2.
\end{equation}
Here, $M$ and $a$ are the Arnowitt-Deser-Misner mass and the angular momentum per unit mass, respectively. Different from that of Kerr-Newman BH, the charge of Kerr-Sen BH is introduced through parameter $b$ in the following way
\begin{equation}
	b=\frac{Q^2}{2M}.
\end{equation}
The Maxwell and dilaton fields are respectively given by
\begin{equation}
	\label{eq:maxwell}
	A_\alpha dx^\alpha=-\frac{Qr}{\rho^2}\left(dt-a\sin^2\theta d\phi\right),
\end{equation}
\begin{equation}
	e^{2\tilde{\Phi}}=\frac{r^2+a^2\cos^2\theta}{\rho^2}.
\end{equation}
The (event and inner) horizons of the Kerr-Sen BH are located at the zeros of $\Delta$,
\begin{equation}
	r_\pm=M-b\pm\sqrt{(M-b)^2-a^2}.
\end{equation}
For extremal BHs,
\begin{equation}\label{extremal BH condition}
	r_{+}=r_{-}=a=M-b,
\end{equation}
which means the following relation should be satisfied for Kerr-Sen BHs,
\begin{equation}\label{relation between b,Q and M}
	b\leq M,\;\;\mathrm{or},\;\;Q\leq\sqrt{2}M.
\end{equation}
In the static limit($a=0$), Kerr-Sen solution reduces to the GMGHS solution \cite{GIBBONS1988741,PhysRevD.43.3140,PhysRevD.88.127901}, while the Kerr solution is recovered in the electrically neutral case ($b=0$).

We shall consider a charged massive scalar test field $\Psi$ in the background of Kerr-Sen  spacetime. The equation of motion of the field  reads
\begin{equation}\label{The Klein-Gordon equation}
	\left[(\nabla^\alpha-iqA^\alpha)(\nabla_\alpha-iqA_\alpha)-\mu^2\right]\Psi=0,
\end{equation}
where $q$ and $\mu$ are respectively the charge and mass of the scalar field and the electromagnetic potential  $A_{\mu}$ of the BH is given by Eq.(\ref{eq:maxwell}).
In order to solve this equation, one can decompose the radial and angular parts of the scalar field  as usual and express $\Psi$ in the form\cite{1.1539899,PhysRevD.92.064022}
\begin{equation}
\Psi_{lm}=e^{-i\omega t+im\phi}R_{lm}(r)S_{lm}(\theta),
\end{equation}
where $S_{lm}(\theta)$ are the spheroidal harmonics which are the solutions of the following differential equation
\begin{equation}\label{spheroidal equations}
		\frac{1}{\sin\theta}\frac{d}{d\theta}\left(\sin\theta\frac{dS_{lm}}{d\theta}\right)+\left[K_{lm}+(\mu^2-\omega^2)a^2\sin^2\theta-\frac{m^2}{\sin^2\theta}\right]S_{lm}=0,
\end{equation}
where, $K_{lm}$ are separation parameters. In the regime $a^2(\mu^2-\omega^2)\lesssim m^2$, one can expand $K_{lm}$ as
\begin{equation}
	K_{lm}=\sum_{k=0}^{\infty}c_k\left[a^2(\mu^2-\omega^2)\right]^k,
\end{equation}
where $c_0=l(l+1)$ and other coefficients $c_k$ can be found in Refs. \cite{Olver:2010:NHMF,NIST:DLMF}. The radial functions $R_{lm}$ are determined by
\begin{equation}\label{Radial equation}
	\Delta\frac{d}{dr}\left(\Delta\frac{dR_{lm}}{dr}\right)+UR_{lm}=0,
\end{equation}
where
\begin{eqnarray}
		U&=&\left[(r^2+2br+a^2)\omega-qQr-ma\right]^2\nonumber\\&&+\Delta\left[2am\omega-K_{lm}-\mu^2(r^2+2br+a^2)\right].
\end{eqnarray}

The radial equation (\ref{Radial equation}) can be also rewritten in the form of a Schr\"{o}dinger-like wave equation
\begin{equation}
\frac{d^2\psi_{lm}}{dr^2}+\left(\omega^2-V(r)\right)\psi_{lm}=0,
\end{equation}
where we have defined $\psi_{lm}\equiv\Delta^{\frac{1}{2}}R_{lm}$ and the effective potential 
\begin{equation}\label{effective_Potential}
V(r)=\omega^2-\frac{U+(M-b)^2-a^2}{\Delta^2}.
\end{equation}
Interestingly, although the effective potential $V$ is different from that in the Kerr-Newman background, they show  the same asymptotic behaviour at infinity 
\begin{equation}\label{effective potential}
V(r)=\mu^2-\frac{2(2M\omega^2-qQ\omega-M\mu^2)}{r}+\mathcal{O}\left(\frac{1}{r^2}\right).
\end{equation}
Therefore, following the arguments of Ref.\cite{PhysRevD.94.064030}, we conclude that the existence condition  for the bound-state resonances of the scalar field in the Kerr-Sen spacetime is the same as Eq.(\ref{bound state condition}).

To solve Eq.(\ref{Radial equation}) numerically,   boundary conditions at spatial infinity and the event horizon $r_+$  should be appropriately specified. 
The asymptotic behaviour of a general solution for Eq.(\ref{Radial equation}) at spatial infinity is given by 
\begin{equation}\label{solution at infinity}
R_{lm}(r\rightarrow \infty)\sim C_1r^{\chi-1}e^{k_\infty r}+C_2r^{-\chi-1}e^{-k_\infty r},
\end{equation}
where
\begin{equation}
k_\infty=\sqrt{\mu^2-\omega^2},\;\;\;\chi=\frac{M(\mu^2-2\omega^2)+qQ\omega}{k_\infty},
\end{equation}
and $\mathrm{Re}(k_\infty)>0$. In the present paper, we shall look for the bound-state resonant modes of the field decaying exponentially at infinity,  so that we  set the coefficient $C_1=0$ in Eq.(\ref{solution at infinity}) \cite{PhysRevD.76.084001,PhysRevD.85.044043,PhysRevD.86.104017,S0217751X13400186,PhysRevD.88.023514}. 

On the other hand,  one usually imposes a purely ingoing boundary condition at the event horizon, \cite{PhysRevD.70.044039}
\begin{equation}\label{horizon behavior}
R_{lm}(r\rightarrow r_{+})\sim e^{-i(\omega-\omega_c)r_*},
\end{equation}
where $r_*$ is the tortoise coordinate, defined by
\begin{equation}
\frac{dr_*}{dr}=\frac{\Delta+2Mr}{\Delta}.
\end{equation}
The critical frequency $\omega_c$ in this situation reads
\begin{equation}\label{def: omega_c}
	\omega_c=m\Omega_H+q\Phi_H,
\end{equation}
where $\Omega_H=\frac{a}{2Mr_+}$ and $\Phi_H=\frac{Q}{2M}$ are, respectively, the angular velocity and the electric potential of the BH's horizon.

In practice, when $\omega=\omega_c$, the near-horizon solution of the radial equation (\ref{Radial equation})  can be expressed as a power series
\begin{equation}\label{series expansion}
R_{lm}=R_0\left(1+\sum_{k=1}^{\infty}R_k(r-r_+)^k\right),
\end{equation}
where $R_0$ is an arbitrary nonzero constant and can be set to $1$ without loss of generality. The coefficients $R_k$ can be found by substituting Eq.(\ref{series expansion}) into Eq.(\ref{Radial equation}) and solving it   order by order in terms of $(r-r_{+})$.   It is worth pointing out that for the case of  extremal BH, the coefficients $R_k$  cannot be solved directly and the power series (\ref{series expansion}) should be replaced by a Frobenius series.

The boundary conditions specified above will pick out a  family of discrete complex frequencies which correspond to the bound-state resonant modes of the scalar field. Generally speaking, these bound-state modes can be divided into three qualitatively distinct types. The first type modes are decaying ones. For such a mode, the real part of the frequency is greater than $\omega_c$ but the imaginary part is negative. The second type will trigger superradiant instability, due to the fact that the real part of the corresponding  frequency is less than $\omega_c$ for superradiance and the imaginary part is positive. The last type is just we are looking for and these resonant modes, dubbed \emph{scalar clouds}, are stationary and in equilibrium with the BH because their frequencies are real and precisely equal to the critical frequency $\omega_c$.

\section{Scanning the parameter space for the existence of the clouds }
\label{Sec: effective potential}

\subsection{Three constraint lines}\label{subsec:threeConstrLines}
Since the scalar clouds are bound states with $\omega=\omega_c$, the critical frequency itself must satisfy the bound-state condition (\ref{bound state condition}), that is
\begin{equation}\label{Eq: cloud_condition}
	f(\mu,q)<\omega_c<\mu.
\end{equation}
Here, without solving the radial equation, we show  that a direct consequence of above condition is that, for given values of $l$ and $m$, both the mass and charge of the scalar cloud  on a given BH are limited in a finite region in the parameter space of the scalar field.\footnote{Throughout the paper, the mass and charge of the scalar cloud  actually refer to the  mass $\mu$ and charge $q$ of the scalar field, respectively.} To this end, we introduce three constraint lines in the $(\mu,q)$ plane which are defined as follows:
\begin{enumerate}[(a)]
	\item $f(\mu,q)=\omega_c$. Using the definition of $f(\mu,q)$, we find that this line is actually a parabola in the $(\mu,q)$ plane \footnote{Note that, in the Kerr-Newman case, the corresponding line is a branch of hyperbolic curve.}, which is given by
	\begin{equation}
		qQ=\frac{2M^2\mu^2-A^2}{A},
	\end{equation}
	where the dimensionless quantity $A$ is defined as
	\begin{equation}\label{def: A}
		A=\frac{ma}{r_+}.
	\end{equation}
	\item $f(\mu,q)=\mu.$ This is a straight line which can be written in a simpler way
	\begin{equation}
		qQ=M\mu.
	\end{equation}
	\item $\mu=\omega_c.$ This is another straight constraint line in the $(\mu,q)$ plane
	\begin{equation}\label{constraint line (c)}
		qQ=2M\mu-\frac{ma}{r_+}.
	\end{equation}
\end{enumerate}
It is found that the above three lines intersect at one point
\begin{equation}
	P_1:(\mu,q)=(\mu_1,q_1)=\left(\frac{ma}{Mr_+},\frac{m a}{Qr_+}\right).
\end{equation}
Besides, lines (a) and (c) also intersect at another point
\begin{equation}
	P_2:(\mu,q)=(\mu_2,q_2)=\left(0,-\frac{ma}{Qr_+}\right).
\end{equation}
Line (a) divides the parameter space into two parts: in one part, the effective potential (\ref{effective_Potential}) forms a well, while in the other part it does not. On the other hand, line (b) is the boundary between the regions in which the bound-state resonances exist or not. At last, line (c) separates the radial function $R_{lm}$ decaying exponentially  at spatial infinity from those growing up. It is worth pointing out that, unlike line (b), the above properties of lines (a) and (c) hold only for the case of $\omega=\omega_c$.

\begin{figure} 
	\centering   
	\subfigure {   
		\includegraphics[width=0.8\textwidth]{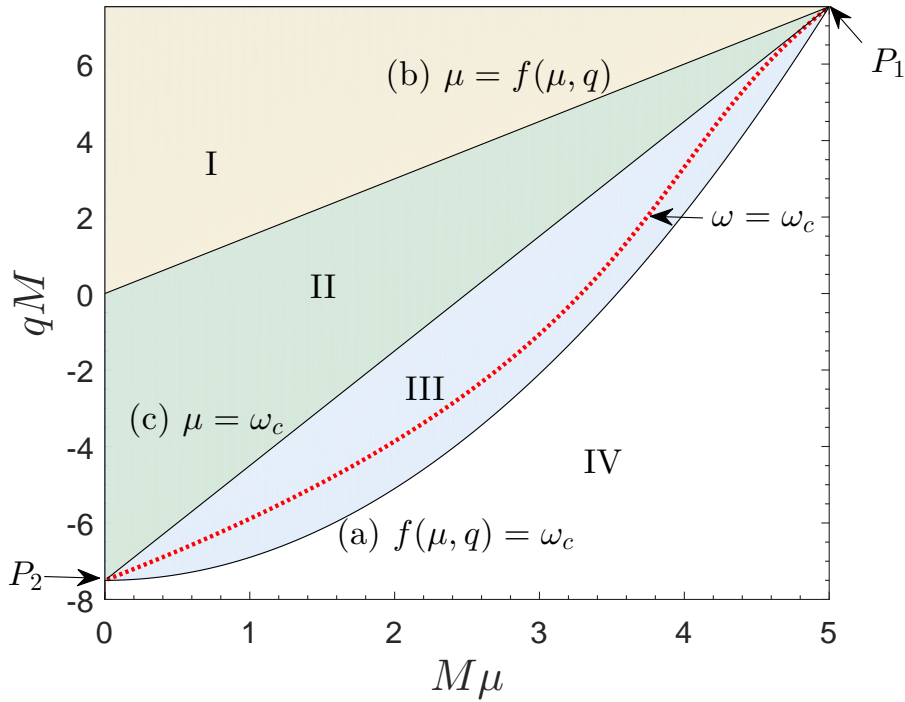} 
	}
	\caption{The parameter space of the scalar field in the Kerr-Sen spacetime is divided by the three constraint lines (a), (b) and (c). For clarity we have set $l=m=5$ and the Kerr-Sen BH to be extremal with the charge $Q=0.6667M$.  The dotted line denotes the existence line  of the cloud with $n=0$ under the same circumstances.}
	\label{fig:2}     
\end{figure}

In  figure \ref{fig:2}, the three constraint lines are plotted in the $(\mu,q)$ plane for the case that $l = m = 5$, $Q = 0.6667M$ and the Kerr-Sen BH is set to be extremal for the sake of clarity.
Clearly, the $(\mu,q)$ plane is divided into four regions by the three constraint lines. In region I, $f(\mu,q)>\mu$, so that no bound-state resonance can be found in this region. On the other hand, although in regions II and IV bound states may exist for some frequencies, it is not difficult to find that  these frequencies are definitely not equal to the critical frequency  $\omega_c$.  Actually, condition (\ref{Eq: cloud_condition}) can only be satisfied in region III. Then, scalar clouds can only exist in this region. Because region III is closed as shown in figure \ref{fig:2}, it is expected that the mass and charge of the clouds are confined in a limited scope. In particular, we notice that  condition (\ref{Eq: cloud_condition}) is marginally met at the points $P_1$ and $P_2$, which suggests that both the maximum and minimum values of the mass and charge of the cloud are determined by the position of the two points.

\subsection{The existence lines}
From the analysis above, it is easy to realize that the clouds exist only for specific values of the parameters.  For instance,  given values of the parameters of BH $\left(M,Q,a\right)$, the harmonic indices  $\left( l,m \right)$ and the mass $\mu$ of the scalar field, we search for values of the field's charge $q$ for which the radial function $R_{lm}$ has a near-horizon behaviour as given by Eq.(\ref{series expansion}) and an exponentially decaying at infinity by numerically integrating Eq.(\ref{Radial equation}). Indeed, we have to assign a set of specific values of $q$, labelled by  the number $n$ of nodes  of the radial function,  in order to obtain a solution with the expected asymptotic behaviour. 
By changing the value of $\mu$ and keeping the values of $n$ and other parameters fixed,  another suitable value of $q$ is acquired. Repeat the above action,  we can take $q$ as a function of $\mu$, 
\begin{equation}
q=q(n,l,m,M,Q,a;\mu).
\end{equation} 
Thus, a curve, called the existence line,
in the parameter space of scalar field (i.e., the two-dimensional $(\mu,q)$ plane) is obtained. Similarly, one can also determine the existence lines in the parameter space of the BH \cite{PhysRevD.90.104024}. In Figure \ref{fig:2}, we also plot the existence line of the clouds ( the dotted line, marked by $\omega=\omega_c$) for the case that the parameters are fixed in the same way as those for the constraint lines. As expected, the existence line is indeed located in region III with the endpoints $P_1$ and $P_2$.

\subsection{No scalar clouds in the GMGHS spacetime}

As a simple application, one can use the analysis method described in subsection \ref{subsec:threeConstrLines} to demonstrate that the GMGHS BH cannot support stationary scalar clouds. For a GMGHS BH, $a=0$, $P_1$ and $P_2$ combine into one point $(0,0)$ in the $(\mu,q)$ plane. Therefore, the existence line is degenerated and no cloud can be stably distributed on the GMGHS BH.

In fact, from Eq.(\ref{def: omega_c}) we have
\begin{equation}
	\omega_c=\frac{qQ}{2M},\;\;\mathrm{for}\;\;a=0.
\end{equation}
However, from the definition of $f(\mu,q)$ in (\ref{bound state condition}),
\begin{equation}
	f(\mu,q)=\frac{qQ}{4M}+\sqrt{\frac{\mu^2}{2}+\frac{q^2Q^2}{16M^2}}>\frac{qQ}{2M},
\end{equation}
which means that no bound-state mode with $\omega=\omega_c$ can be found in this case. Thus, we conclude that GMGHS BH cannot support linear charged massive scalar clouds\footnote{The author of \cite{PhysRevD.94.085007} drew the conclusion that the GMGHS BH can support scalar clouds, based only on the fact that $\tilde{\alpha}\neq 0$ in the case of $a=0$, where $\tilde{\alpha}$ is defined by Eq.(55) in \cite{PhysRevD.94.085007}. But just as pointed out in \cite{PhysRevD.94.085007}, $\tilde{\alpha} >0 $ is required for the existence of the cloud. Indeed, we find that $\tilde{\alpha} <0 $ when $a=0$.}. This result is consistent with the stability analysis in Ref.\cite{PhysRevD.88.127901}. In fact, it was shown in Ref.\cite{Li:2015bfa} that scalar clouds can only form on the GMGHS background in the presence of a reflecting mirror surrounding the BH.

\section{The existence lines for scalar clouds}\label{Sec: existence lines}

In this section, we give our numerical results and exhibit the existence lines of the scalar clouds in the parameter space of both the Kerr-Sen BH and the scalar field.

\subsection{The existence lines in the parameter space of the Kerr-Sen BH}\label{subSec: A}

\begin{figure}
	\centering    
	\subfigure {   
		\includegraphics[width=0.8\textwidth]{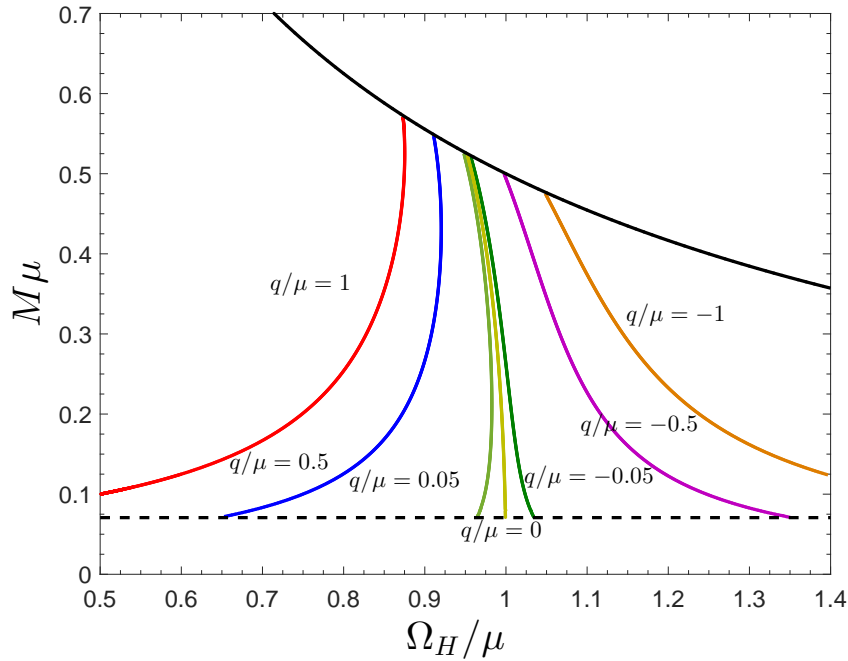} 
	}
	\caption{The existence lines of scalar clouds on the Kerr-Sen BH with the charge $\mu Q=0.1$, for different values of the charge-to-mass ratio $q/\mu$ of the field.  Here, we fix $n=0$ and $l=m=1$. The dashed black line denotes $M\mu=Q\mu/\sqrt{2}$, while the solid black one corresponds to the extremal BH.}
	\label{fig:1}     
\end{figure}

Figure \ref{fig:1} shows the existence lines for scalar clouds in the parameter space of Kerr-Sen BH for various values of the charge-to-mass ratio of the scalar field. The overall trend of these existence lines is similar to that in the Kerr-Newman case. That is,  clouds with charges of the same (opposite) sign as the BH occur for smaller (larger) angular velocities \cite{PhysRevD.90.104024}. This result is consistent with our intuition.

From figure \ref{fig:1}, we further observe that the existence lines do not reach the line $M =0$, but end at the points where $M$ takes some nonzero value. This feature is also found in the Kerr-Newman case. However, the minimum value of $ M$ is $ Q/\sqrt{2}$, while it is $ M= Q$ for the clouds on a Kerr-Newman BH \cite{PhysRevD.90.104024}. This difference comes from the different conditions for the absence of naked singularity in Kerr-Newman and Kerr-Sen spacetimes: For a Kerr-Newman spacetime, the condition is $a^2+Q^2\leq M^2$, while for a Kerr-Sen spacetime, it is $2Ma+Q^2\leq 2M^2$.

\subsection{The existence lines in the $(\mu,q)$ plane}
\begin{figure}
	\centering 
    \includegraphics[width=0.8\textwidth]{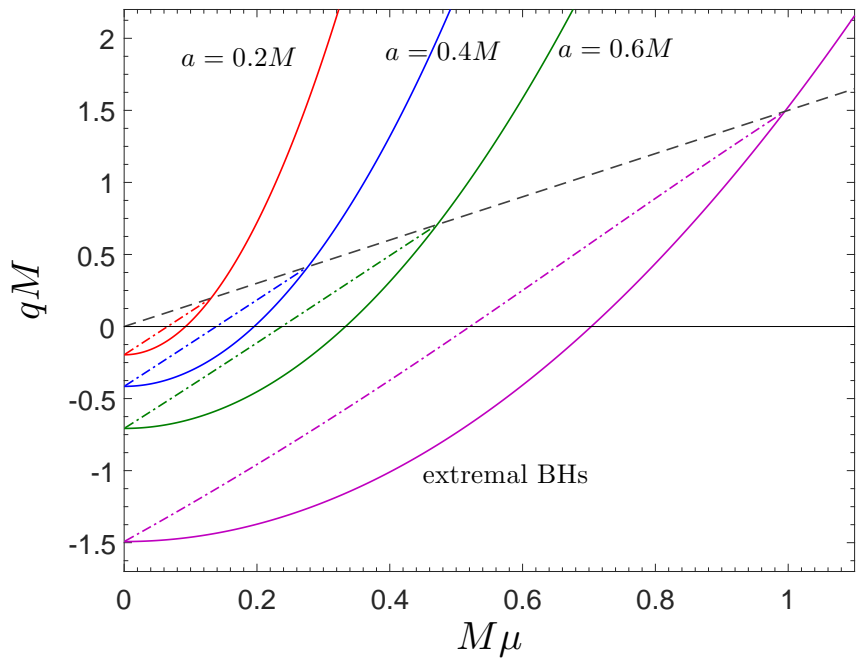} 
	\caption{ The existence lines for scalar clouds (dot-dashed) in the $(\mu,q)$ plane, with $l=m=1$, for different values of $a$ of the Kerr-Sen BH. Here, we fix $Q=0.6667M$ and also plot corresponding constraint lines (a) (solid) and (b) (dashed). }
	\label{fig:3}     
\end{figure}

\begin{figure}
	\centering    
    \includegraphics[width=0.8\textwidth]{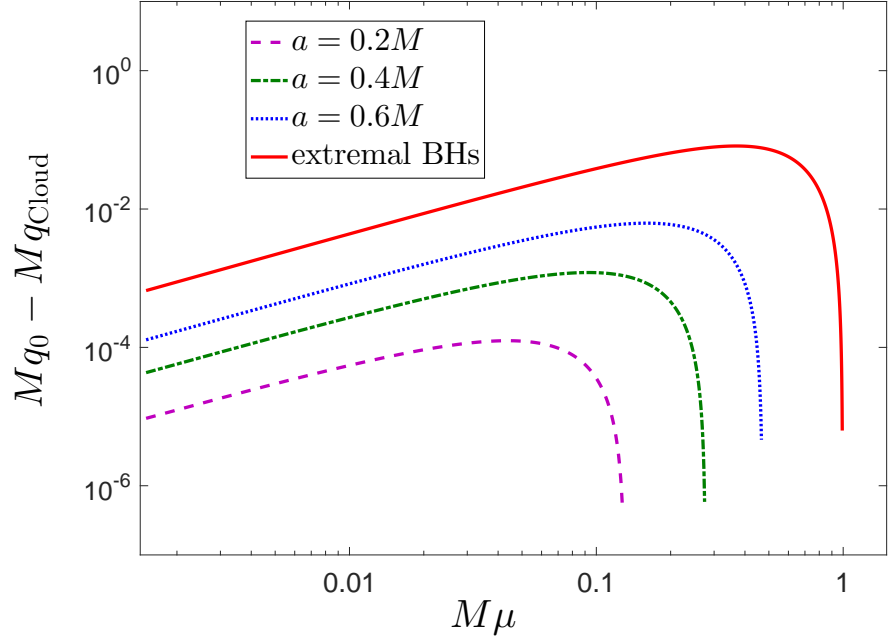} 
	\caption{The deviation of the existence line from the corresponding constraint line (c) between intersections $P_1$ and $P_2$ for different values of $a$ of the Kerr-Sen BH.}
	 \label{fig:3b}       
\end{figure}

In figure  \ref{fig:3}, we fix $Q=0.6667M$ and plot the existence lines for the clouds with $l=m=1$ for various values of $a$ of the Kerr-Sen BH. Here, we also add corresponding constraint lines (a) and (b) to show that the endpoints of the existence lines are indeed the two intersections $P_1$ and $P_2$ for given values of $a$. Interestingly, the existence lines are almost straight lines between the two intersections for different values of $a$. In other words, the existence line is very close to the corresponding constraint line (c).  We further show the deviation of the existence line from corresponding line (c) for various values of $a$ in figure \ref{fig:3b}, in which the vertical axis represents the difference $Mq_0-Mq_{\mathrm{cloud}}$, where $Mq_0$ is evaluated from Eq.(\ref{constraint line (c)}),  but  $Mq_{\mathrm{cloud}}$ is directly given  by the charge of the scalar field.  One find that the faster the Kerr-Sen BH rotates, the greater the deviation will be. Consequently, when $a$ is small enough, the existence lines of the clouds can be well approximated by Eq.(\ref{constraint line (c)}), i.e.,
\begin{equation}
	Qq_{\mathrm{cloud}}\approx2M\mu_{\mathrm{cloud}}-\frac{m a}{r_+}.
\end{equation}

\begin{figure}
	\centering    
	\subfigure {    
		\includegraphics[width=0.8\textwidth]{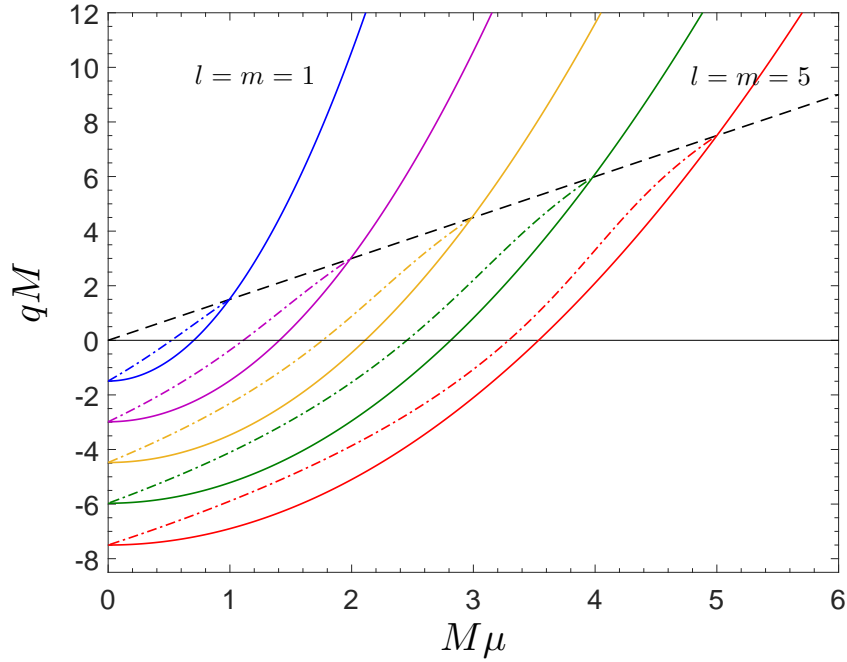} 
	}
	\caption{The existence lines (dot-dashed) for scalar clouds on the extremal Kerr-Sen BH with $Q=0.6667M$  for the values of harmonic indices $l=m=1,2,3,4$ and $5$. Here, we also include the constraint lines (a) (solid lines, depending on the value of $m$) and (b) (black dashed line).}
	\label{fig:4}     
\end{figure}

\begin{figure*} 
	\centering   
	\subfigure {   
		\includegraphics[width=0.8\textwidth]{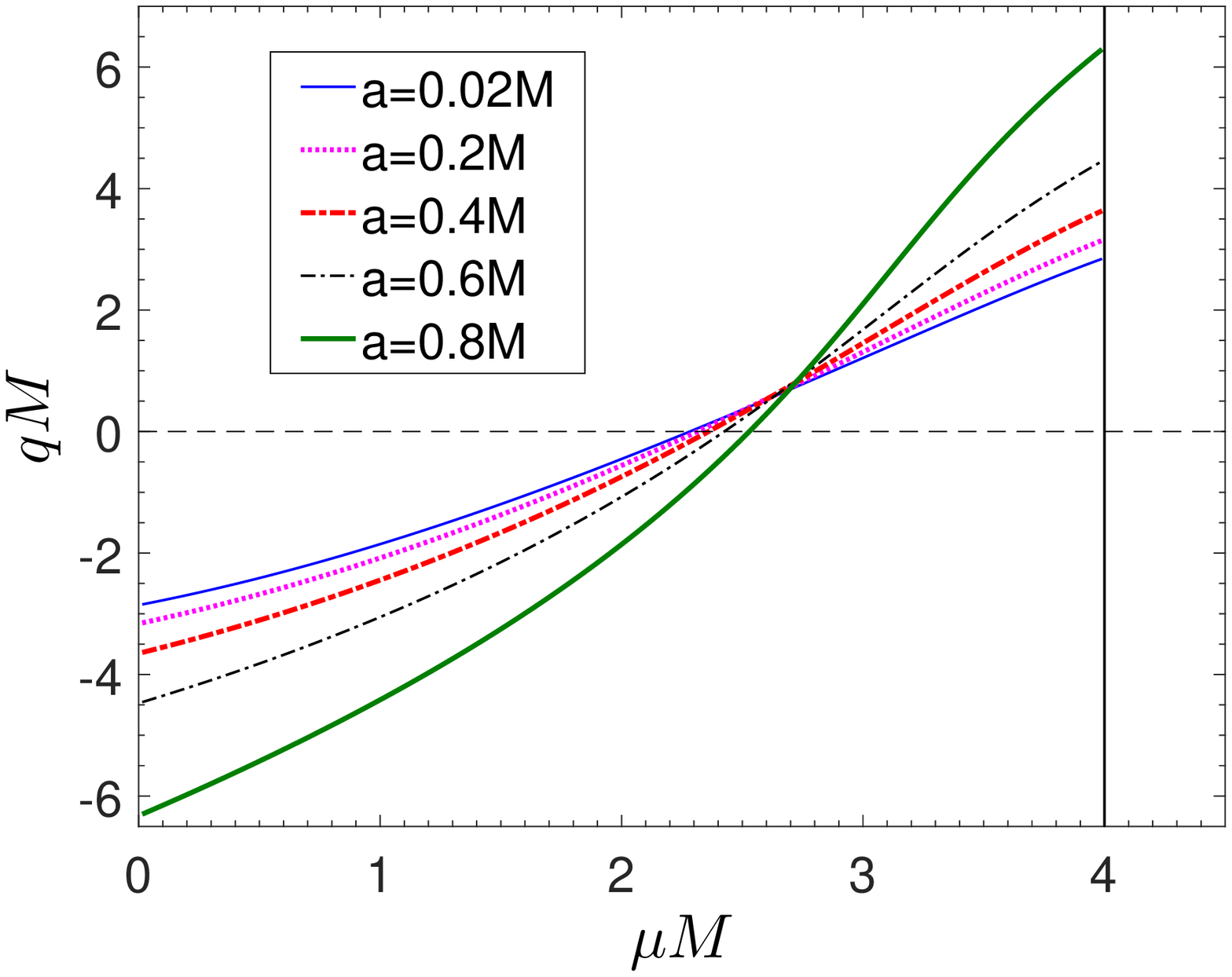} 
	}
	\caption{The existence lines for scalar clouds in the $(\mu,q)$ plane with $l=m=4$ for various values of $a$ of the extremal Kerr-Sen BH. In order to illustrate that the maximum values of the mass of the scalar clouds are given by $M\mu=m$ as is presented in Eq.(\ref{maximum values of mu and q}), we also plot the vertical line ($M\mu=4$).}
	\label{fig:5}     
\end{figure*}

\begin{figure*}\centering
	\subfigure {    
		\includegraphics[width=0.8\textwidth]{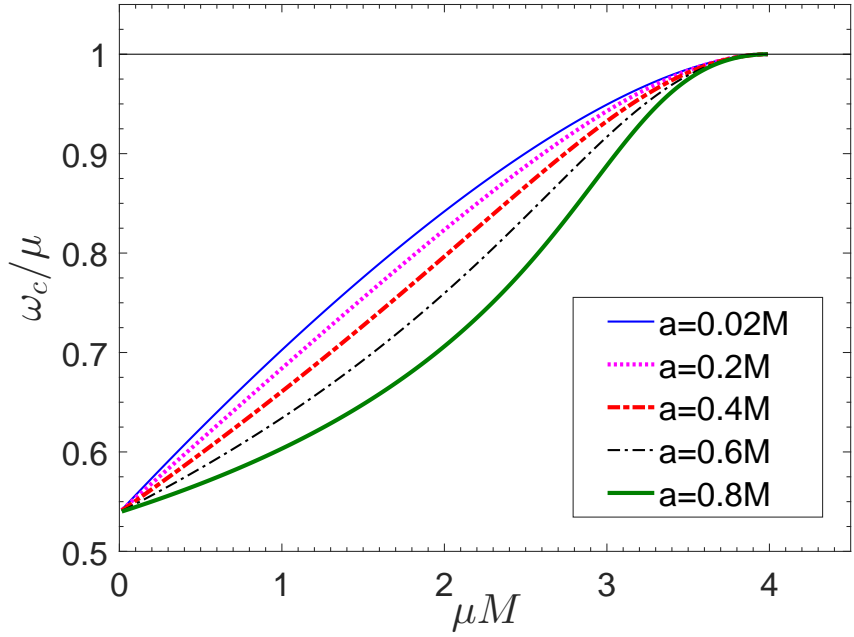} 
	}
	\caption{The values of the ratio $\omega_c/\mu$ along the existence lines in Figure \ref{fig:5} for corresponding values of $a$. The horizontal line is to show that when the dimensionless mass $M\mu$ approaches its maximum value, $\omega_c/\mu\rightarrow1$.}
	\label{fig:6}
\end{figure*}

The existence lines (dot-dashed) for the clouds in extremal Kerr-Sen background are plotted in figure \ref{fig:4} where the charge of the BH is fixed and the harmonic indices take different values, and in figure \ref{fig:5} where  $l=m=4$ is fixed and several different  values of $a$ are selected. In addition, we also keep track of the ratio $\omega_c/\mu$ along the existence lines in figure \ref{fig:5} and exhibit the result in figure \ref{fig:6}. From figure \ref{fig:6},  it is found that along the existence line $\omega_c/\mu$ is always less than $1$ and $\omega_c/\mu\rightarrow 1$ as  the mass of the cloud approaches to its maximum value. This result is easy to understand because, as is figured out in section \ref{Sec: effective potential}, the existence line of the scalar cloud is located below the constraint line (c) and the  line tends to be tangent to the existence line near the intersection $P_1$.

For extremal Kerr-Sen BHs, there exists another interesting feature: the maximum values of $\mu$ and $q$ for the scalar cloud are independent of $a$ which governs the angular velocity of the BH, and more specifically
\begin{equation}\label{maximum values of mu and q}
	M\mu_{\mathrm{max}}=Qq_{\mathrm{max}}=m.
\end{equation}
It should be pointed out that the results are radically different from those in the case of extremal Kerr-Newman BH. As was shown in \cite{PhysRevD.94.064030}, the maximum values of the cloud mass and charge in the Kerr-Newman spacetime are given by
\begin{equation}
	\mu_{\mathrm{max}}=\frac{ma}{Mr_+-Q^2},\;\;q_{\mathrm{max}}=\frac{mMa}{Q(Mr_+-Q^2)}.
\end{equation}
If the BH is extremal, $r_+=M,\;\;a^2+Q^2=M^2$, then, one finds
\begin{equation}
	a\mu_{\mathrm{max}}=m,\;\;aQq_{\mathrm{max}}=mM.
\end{equation}
However, for a given Kerr-Sen BH, the mass $\mu$ and charge $q$ of the cloud take their maximum values at $P_1$ as
\begin{equation}
	M\mu_{\mathrm{max}}=Qq_{\mathrm{max}}=\frac{m a}{r_+}.
\end{equation} 
When the Kerr-Sen BH is extremal,  $r_+=a$ from Eq.(\ref{extremal BH condition}), then Eq.(\ref{maximum values of mu and q}) is recovered.

\section{Conclusions and discussions}\label{Sec: conclusion}

In summary, we have performed a numerical analysis of scalar clouds around Kerr-Sen BHs. These stationary bound-state solutions are characterized by the existence lines in the parameter space for both the Kerr-Sen BH and the scalar field. Similar to the Kerr-Newman case, the cloud existence lines in the parameter space of the Kerr-Sen BH are confined in a finite region. However, the minimum value of the mass of the BH are different in the two cases. 

Furthermore, for a given Kerr-Sen BH, the parameter space of the scalar field is divided into four regions by three simple constraint lines when the harmonic indices are fixed and the existence line can be only located in one of the four regions. Because the region is closed, it is concluded that  the mass and charge of the clouds are confined in a limited scope.  

In addition, we also find that the maximum values of mass and charge for scalar clouds around extremal Kerr-Sen BHs are independent of $a$ of the BH, whereas the corresponding maxima for a Kerr-Newman BH do depend on $a$. The difference originates from the basic fact that the horizon radius $r_{+}=a$ for an extremal Kerr-Sen BH, while $r_{+}=M$ in the Kerr-Newman case.
This reflects a difference between low-energy heterotic string theory and Einstein-Maxwell theory that electric charges in the two theories exert their influences  on the metric of the spacetime in different ways.

It is important to note that no stationary scalar clouds can be supported on the Kerr-Sen BH in the static limit (i.e., GMGHS BH), since no bound states with $\omega=\omega_c$ can be found in this situation. However, the so-called "marginal clouds" with $M\mu=qQ$, may be able to exist \cite{PhysRevD.90.064004,Degollado2013}.

Let us mention that the present work is limited to charged massive scalar fields minimally coupled to the Kerr-Sen BHs.  For scalar field with a suitable self-interacting potential \cite{PhysRevD.92.084059,Herdeiro2014302} and Proca field \cite{PhysRevD.86.104017,PhysRevLett.109.131102} in the background of a Kerr-Sen BH, stationary bound states may exist even if the equation of motion for the test field is not separable. An appealing question is whether or not the analogue bounds  on the parameters exist in these scenarios. Finally, it would be interesting to generalize our results for the linear charged scalar clouds to the fully nonlinear regime of charged scalar hairs \cite{PhysRevLett.112.221101,Delgado2016234} in the Kerr-Sen spacetime. These topics deserve new works in the future.

\ack
This work is supported in part by National Natural Science Foundation of China under Grant No. 11275128, Science and Technology Commission of Shanghai Municipality under Grant No. 12ZR1421700, and the Program of Shanghai Normal University.

\section*{References}
\bibliographystyle{unsrt}       
\bibliography{Ref}

\end{document}